# Big Data, Socio-Psychological Theory, Algorithmic Text Analysis, and Predicting the Michigan Consumer Sentiment Index

Rickard Nyman[*], Paul Ormerod

Centre for the Study of Decision Making Under Uncertainty,
University College London and Diphrontis Analytics Ltd., London

May 2014

- Corresponding author  rickardnyman@gmail.com

**Abstract**

We describe an exercise of predicting the Michigan Consumer Sentiment Index, a widely used indicator of the state of confidence in the US economy. We carry out the exercise from a pure *ex ante* perspective. We use the methodology of algorithmic text analysis of an archive of brokers' reports over the period June 2010 through June 2013. The search is directed by the social-psychological theory of agent behaviour, namely conviction narrative theory.

We compare one month ahead forecasts generated this way over a 15 month period with the forecasts reported for the consensus predictions of Wall Street economists. The former give much more accurate predictions, getting the direction of change correct on 12 of the 15 occasions compared to only 7 for the consensus predictions. We show that the approach retains significant predictive power even over a four month ahead horizon.

1. **Introduction**

The development of Big Data appears to provide many opportunities for discovering knowledge in hitherto unconventional ways. However, it is essential to proceed with caution. An enormous amount of data has become available, and as a result there will be many apparently significant correlations awaiting discovery. However, as Silver points out in his best-selling book *The Signal and the Noise*, many of these will be spurious and almost entirely dependent upon the particular sample which is used to estimate them.

In order to interpret correctly any relationships which are found in Big Data, it is essential to view the results from the perspective of a soundly based behavioural theory. In terms of frequentist statistical theory, a correlation between any two factors is deemed to be significant if, to use natural rather than scientific language, the chance of observing the correlation is less than 1 in 20. Given the immense amount of data which is now available, and the ease of processing it, it is clearly very easy to obtain large numbers of correlations which meet this criterion.

However, results obtained through data mining in this way can only be interpreted *ex post*. Almost anything can be rationalised in this way. Without theoretical guidance, there is no way of knowing in advance even what the sign of any correlation ought to be. In this paper, we describe an exercise of predicting the Michigan Consumer Sentiment Index, a widely used indicator of the state of confidence in the US economy. We carry out the exercise from a pure *ex ante* perspective. We use the methodology of algorithmic text analysis of an archive of brokers' reports over the period June 2010 through June 2013.

A key point here is that our text analysis is guided completely by the social-psychological theory of conviction narratives (Chong and Tuckett, 2013; Tuckett, Smith and Nyman, 2013). The theory starts from the proposition agents may be more or less emotionally convinced that the information they have available provides the grounds to make confident decisions.

Indicators of the emotional conviction in narratives are then used in conjunction with standard algorithmic text search methodologies to filter and extract shifts in confidence from the database.

Section 2 describes the Michigan data and the accuracy of the forecasts of the index made by a survey of economists polled by Reuters. Section 3 sets out the methodology we employ, and section 4 compares the forecasts with those of the economic consensus.

## 2       The Michigan index and the consensus forecasting record

The Michigan Consumer Sentiment Index has long been the industry leading measure of consumer confidence and consumer expectations. Survey results are released twice each month at 10.00 a.m. Eastern Time: preliminary estimates usually (variations occur during the winter season) on the second Friday of each month, and final results on the fourth Friday.

There is a very high correlation of 0.966 between the preliminary and final index. The correlation between the change in the preliminary index in any given month from the final in the previous month, and change in the final itself remains high, at 0.905. When the economists are polled by Reuters, the preliminary estimate for the month has already been published. So the task of predicting the final value of the index is considerable simplified by having this information.

The real question is what the change will be in the preliminary index from the level of the final index in the previous month. Given the preliminary index and its high correlation, whether in levels or differences, with the final, the forecasting task is fairly straightforward.

In terms of the change in the preliminary index from the level of the final index in the previous month, the forecasting accuracy of the consensus is considerably less impressive, indeed it might even be described as poor. As a matter of descriptive convenience, we will now refer to this change in the preliminary index from the level of the final index in the previous month as DIFFPRELIM.

The consensus forecasts only predict the sign of DIFFPRELIM correctly on 7 out of the 15 occasions. This is no better than a purely random guess. A linear regression of DIFFPRELIM on the change in the preliminary on the previous final predicted by the consensus forecast confirms the poor record (which we describe as DIFFCONSENSUS)[1].

Over the period May 2012 through July 2013,

(1)        *DIFFPRELIM = -1.293 + 1.972*DIFFCONSENSUS*

*(1.084)   (1.178)*

---
[1] The autocorrelation function of each variable contains no lags which are significantly different from zero, so the two variables have the same order of integration

*Residual standard error: 4.045    Adjusted R-squared 0.114*

*F-statistic: 2.804 on 1 and 13 degrees of freedom, p-value: 0.118*

*The figures in brackets are the estimated standard errors of the coefficients.*

The explanatory power of the equation is very low. In fact, it is essentially not significantly different from zero. In other words, the consensus forecasts have very little value in terms of predicting the change in the preliminary index from the level of the final index in the previous month.

We now move to consider whether the algorithmic text based approach, filtered through the lens of conviction narrative theory, can do any better.

### 3. Algorithmic text analysis based on conviction narrative theory

We analyse an archive of 14 brokers from June 2010 through June 2013 consisting of documents of a primarily global economic focus. The archive consists of approximately 111 documents per month. The documents are very long (up to 50 pages in some cases), and so we pick up on a large number of words. In total we arrive at 37 monthly data points.

The approach we use here is simply one particular application of a methodology which has been developed to analyse *any* textual data base.

A detailed description of the approach, including the algorithm used for text analysis, is available in Tuckett et al. (2013). Here, we provide a summary.

The social-psychological theory of conviction narratives (Chong and Tuckett, 2013; Tuckett, Smith and Nyman, 2013) starts from the proposition agents may be more or less emotionally convinced that the information they have available provides the grounds to make confident decisions. A conviction narrative combines reasons for and against action into order in such a way that action is supported. Agents faced with uncertainty can then feel persuaded to act and to stay acting, while they try to interpret signs and signals in the world and wait to see how their decisions are turning out.

Indicators of the emotional conviction in narratives are then used in conjunction with standard algorithmic text search methodologies to filter and extract shifts from any given text data base.

For any given text data base, we compute two emotional summary statistics, one for *excitement* (the attractor) and one for *anxiety* (the repellor), by applying a simple word count methodology. Two sets of emotion words, each of size approximately 150, indicative of the relevant emotions have been defined. The lists proved useful in other studies (Tuckett, Smith and Nyman, 2013, Tuckett et al. 2013) and have been validated in a laboratory setting (Strauss, 2013).

We construct from this a single variable which we use for analytical purposes, defined as the difference between the frequency of excitement words and the frequency of anxiety words, normalized by dividing by the total number of characters[2] in the data base. For descriptive convenience we describe this as BROKER.

### 4. The results

The approach we use is as follows. We estimate a linear regression of DIFFPRELIM on DIFFBROKER in the previous month, where the latter is the change in BROKER.

The time stamp of the data is important to explain. It is crucial to understanding the significance of the results.

Initially, we estimate the regression using the data on DIFFPRELIM from August 2010 through April 2012. The first observation in this sample is the value of the preliminary index in August 2010 minus the value of the final observation in July 2010. The corresponding data point for the series BROKER is the change in the value of BROKER between July and June 2010. In other words, we regress DIFFPRELIM on information which would have been available at the end of the *previous* month to which DIFFPRELIM relates.

Using data on DIFFPRELIM from August 2010 through April 2012 and data on DIFFBROKER from July 2010 through March 2012, we obtain[3]:

(2) DIFFPRELIM = -0.877 + 0.681*DIFFBROKER

     (0.766)   (0.261)

*Residual standard error: 3.486    Adjusted R-squared 0.225*

*F-statistic: 6.815 on 1 and 19 degrees of freedom, p-value: 0.017*

*DW = 1.92; Ramsey F (3,30) = 0.69; W = 0.96*

*The figures in brackets are the estimated standard errors of the coefficients; DW is the Durbin-Watson statistic for first order autocorrelation; Ramsey is the Ramsey RESET specification test and W is the Shapiro-Wilk test for normality of the residuals*

The equation is well-specified.

Using the full sample, in other words DIFFPRELIM from August 2010 through July 2013 and data on DIFFBROKER from July 2010 through June 2013, we have:

(3) DIFFPRELIM = -0.787 + 0.788*DIFFBROKER

---

[2] In general, the total number of words or documents is the divisor, but in this particular instance some of the documents contain tables and others do not, so that the total number of characters is more appropriate

[3] Again, the autocorrelation function of each variable contains no lags which are significantly different from zero, so the two variables have the same order of integration

*(0.536)   (0.170)*

*Residual standard error: 3.212   Adjusted R-squared 0.369*

*F-statistic: 21.49 on 1 and 34 degrees of freedom, p-value: 0.00005*

*The figures in brackets are the estimated standard errors of the coefficients*

*DW = 1.97; Ramsey F (3,15) = 0.81; W = 0.94*

Again, the equation is well-specified.

To generate forecasts of the preliminary estimates of the Michigan index for May 2012, we use the coefficients in equation (2) above, and the data for DIFFBROKER in April 2012. In other words, to predict the May value of the index, we use information which was available at the end of April.

We then repeat the analysis, moving the sample forward one month at a time, until we predict the index in July 2013 using the equation estimated with DIFFPRELIM from August 2010 through June 2013 and DIFFBROKER July 2010 through May 2013. The prediction for July 2013 uses the value of DIFFBROKER in June 2013. Again, to emphasise, when making the prediction we only use information which was available at the previous month. This replicates as far as possible an *ex ante* forecasting situation.

We also emphasise that the text analysis was only carried out once. In other words, we applied our general methodology to this particular data base and used the results to make predictions, as described above. We did not do repeated searches of the data base, using for example only sub-sets of the complete set of words which represent excitement and anxiety, or giving words different weights in order to improve the forecast performance. *Ex post,* it would almost certainly be possible to achieve an apparent improvement in 'forecast' performance by carrying out such procedures, but as a way of replicating an *ex ante* forecasting situation, it would be wholly invalid.

Further, we specified the very simple functional form in equations (2) and (3) and then carried out the regressions. We did not modify this in any way in order for the equations to perform better on statistical tests of validation. The test statistics reported with the equations therefore satisfy completely the requirements of statistical theory and their power can be relied upon. We make this point because many regressions, especially on time series data, reported in the academic econometric literature, appear to satisfy an impressive battery of specification tests. But usually this is only achieved by modifying the specification of the equation, either in terms of explanatory variables or in terms of functional form, in order that the equation does in fact satisfy such tests. But in these circumstances, the true power of the tests is in general unknown, except that it is less than that suggested by statistical theory.

To recap, the consensus forecasts made by economists over the period May 2012 through July 2013 only get the sign of the change correct on 7 out of 15 equations, and a regression of the actual value of DIFFPRELIM on the changes implied by the consensus forecasts has effectively zero statistical power. This is the benchmark against which we judge our predictions.

Our methodology captures the correct value of the sign of DIFFPRELIM on 12 out of 15 occasions.

The regression comparable to (1) using the BROKER data is as follows:

(4)     DIFFPRELIM( in month t) = 0.347  +  1.219*DIFFBROKER( in month t-1)

                              (0.852)    (0.323)

*Residual standard error: 3.081    Adjusted R-squared 0.486*

*F-statistic: 14.24 on 1 and 13 degrees of freedom, p-value: 0.0023*

*The figures in brackets are the estimated standard errors of the coefficients.*

It is apparent that, whilst the equation is not perfect, it has genuine power and is very much better than equation (1). The predictions are unbiased, given that the intercept is not significantly different from zero and the coefficient on the explanatory variable is not significantly different from one.

The difference in the forecasting performance of the CONSENSUS and BROKER data is seen very clearly in Figures 1a and 1b which plot the difference between the preliminary estimate and the final value in the previous month (DIFFPRELIM) and the prediction given by the consensus and broker approaches, as described above

**Figure 1a**

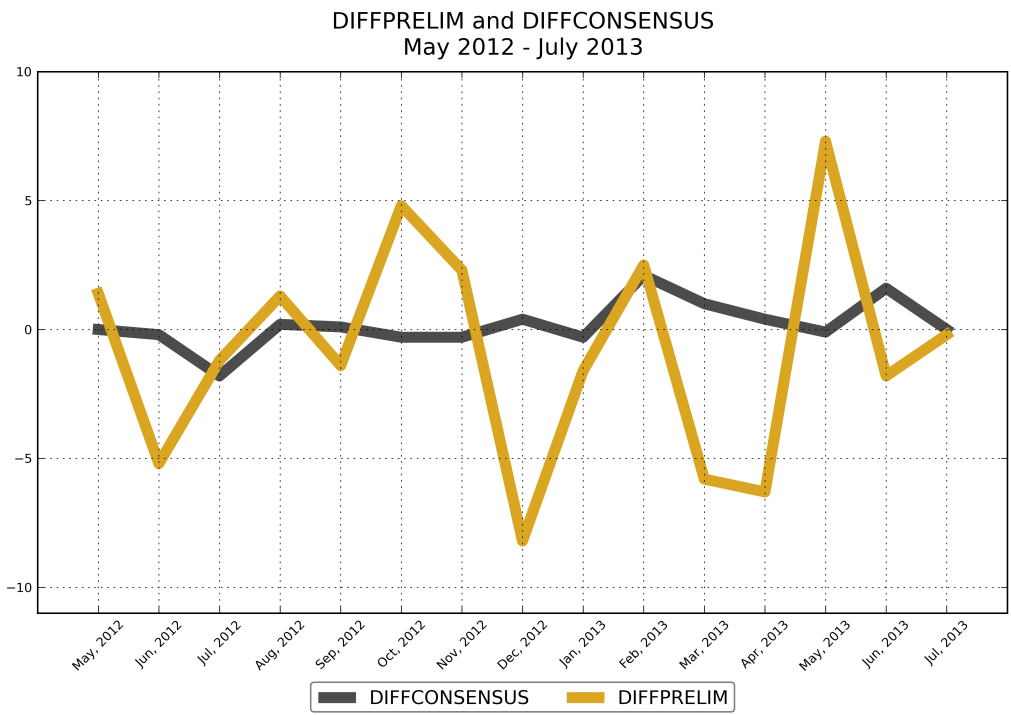

**Figure 1b**

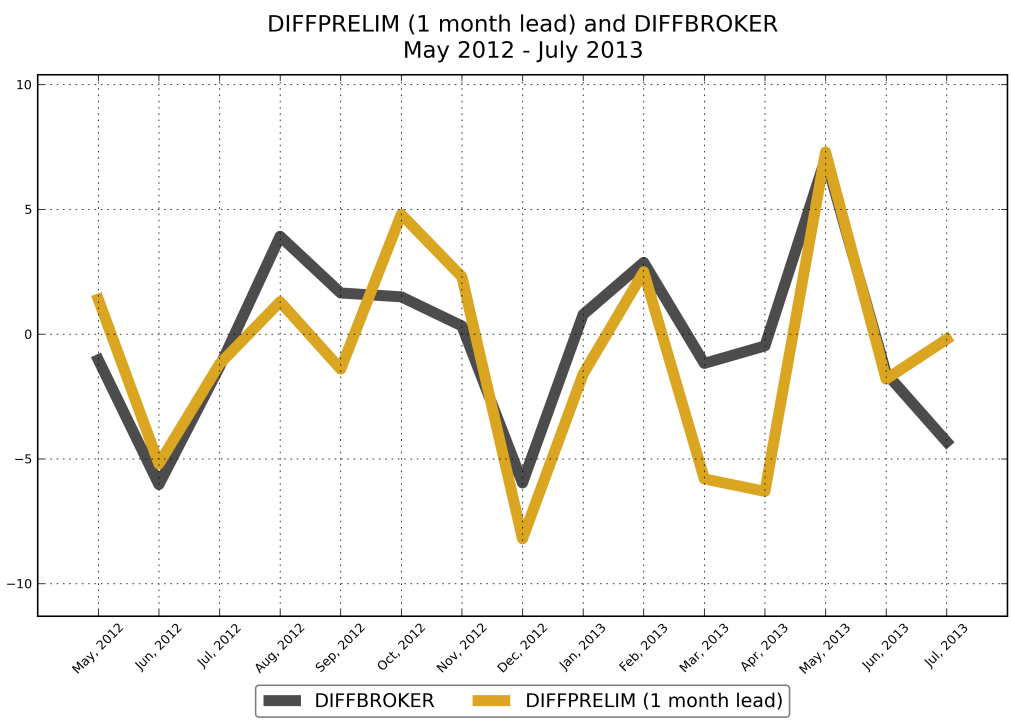

We also examined the ability of the BROKER series to predict the preliminary value of the MCI further ahead than the immediate next month. We consider 2, 3 and 4 months ahead. So, using information available at the end of April 2012, for example, the prediction 2 months ahead is for the change in the preliminary value of the MCI in June 2012 on the final value in April 2012. Similarly, the 3 months ahead is the change in the preliminary value of the MCI in July 2012 on the final value in April 2012, and the 4 month ahead is the change in the preliminary value of the MCI in August 2012 on the final value in April 2012.

Even 4 months ahead, there is some predictive power in the BROKER data, although the performance deteriorates the further ahead the prediction is made, as one would expect. In terms of the correct prediction of the sign of the change in the preliminary index, defined as in the paragraph immediately above, for the 2 month ahead it is 11/15, for the 3 month 8/15 and for the 4 month 7/15.

5. **Discussion and Conclusion**

The Michigan Consumer Sentiment Index is important not only in its own right as an indicator of the current state of consumer confidence in America, but it is also the focus of many trades on financial markets. Economists make predictions of this index, month by month, and their views are polled by Reuters and the consensus is published. The final value of the index for any given month is published essentially at the end of the month, but a preliminary estimate is also published in the first half of the month. This preliminary estimate is very strongly correlated with the final value, and is available to the economists when they make their predictions.

The real challenge, is therefore, to predict not the final, but the preliminary value of the index. More specifically, the challenge is to predict the change in the preliminary estimate from the final value of the previous month. The performance of the economic consensus forecasts of this change over the 15 months from May 2012 through July 2013 is poor. Even the sign of the change is correctly predicted on only 7 out of the 15 occasions, no better than a random guess. A regression of the actual change on the predictive change has essentially no predictive power.

The approach we have presented, grounded in the social-psychological theory of conviction narratives and using directed algorithmic text analysis with a database of brokers' reports generates a time series which indicates the net level of excitement minus anxiety found in the reports.

We replicate as far as possible a genuine ex ante forecasting situation over the same 15 months from May 2012 through July 2013. These predictions give the correct sign on 12 out of the 15 occasions, and have significant explanatory power. The methodology can readily be applied to other text databases in the same or other forecasting contexts. It can undoubtedly be refined. For example, all documents are given equal weight in our analysis, even though in practice some may be more influential than others.